\documentstyle[12pt]{article}
\def\be{\begin{equation}}
\def\ee{\end{equation}}
\def\bea{\begin{eqnarray}}
\def\eea{\end{eqnarray}}

\begin{document}

\begin{center}
{\Large{\bf $SL(2;R)$ Duality of the Noncommutative DBI Lagrangian}}

\vskip .5cm
{\large Davoud Kamani}
\vskip .1cm
 {\it Institute for Studies in Theoretical Physics and
Mathematics (IPM)
\\  P.O.Box: 19395-5531, Tehran, Iran}\\
{\sl e-mail: kamani@theory.ipm.ac.ir}
\\
\end{center}

\begin{abstract}

We study the action of the $SL(2;R)$ group on the noncommutative DBI
Lagrangian. The
symmetry conditions of this theory under the above group will be obtained. 
These conditions determine the extra $U(1)$ gauge field.
By introducing some consistent
relations we observe that the noncommutative (or ordinary) DBI
Lagrangian and its
$SL(2;R)$ dual theory are dual of each other. 
Therefore, we find some $SL(2;R)$ invariant equations. In this case 
the noncommutativity parameter, its $T$-dual and its $SL(2;R)$ 
dual versions are expressed in terms of each other. 
Furthermore, we show that on the effective variables, $T$-duality and 
$SL(2;R)$ duality do not commute. We also study the effects of the
$SL(2;R)$ group on the noncommutative Chern-Simons action.

\end{abstract}
\newpage
\section{Introduction}
$SL(2;R)$ duality generalizes strong-weak coupling duality. 
There is an $SL(2;R)$
symmetry manifest in the low energy action, which is broken down to $SL(2;Z)$
in string theory. Also there is considerable evidence in favor of this 
duality being an exact symmetry of the full string theory \cite{1,2,3}. 
In fact, the $SL(2;R)$ group and its subgroup $SL(2;Z)$ act as symmetry
groups of many theories \cite{4,5,6}. Among these theories, the noncommutative
theories and the Dirac-Born-Infeld (DBI) theory are more important,
for example see the Refs. \cite{6,7}.

Consider the $SL(2;R)$
symmetry of the type IIB superstring theory \cite{1,2,3}.
In the type IIB theory the R-R zero-form $\chi$
and the dilaton $\phi$ of the NS-NS sector define a complex variable
$\lambda = \chi + ie^{-\phi}$. Under the $SL(2;R)$ duality this variable
and also the NS-NS and R-R two-forms $B_{\mu\nu}$ and $C_{\mu\nu}$
transform as in the following
\bea
&~&\lambda \rightarrow {\tilde \lambda}= \frac{a \lambda+b}
{c \lambda+d}, 
\nonumber\\
&~&\left( \begin{array}{c} B_{\mu\nu} \\ C_{\mu\nu} \end{array} \right) 
\rightarrow
\left( \begin{array}{c} {\tilde B}_{\mu\nu} \\ {\tilde C}_{\mu\nu} 
\end{array} \right) = (\Lambda^T)^{-1} 
\left( \begin{array}{c} B_{\mu\nu} \\ C_{\mu\nu} \end{array} \right)
\;\;,\;\;\;\;\;
\Lambda = \left( \begin{array}{cc} a & b\\ 
c & d
\end{array} \right)
\in SL(2;R).
\eea
In addition, the Einstein metric 
$g^{(E)}_{\mu\nu} =e^{-\phi/2}g_{\mu\nu}$ remains invariant. Therefore,
the string coupling constant $g_s = e^\phi$ and the string metric
$g_{\mu\nu}$ transform as follows
\bea
g_s \rightarrow {\tilde g}_s = \eta^2 g_s
\;\;,\;\;\;\;g_{\mu\nu} \rightarrow {\tilde g}_{\mu\nu}=\eta g_{\mu\nu}
\;\;\;,\;\;\;\eta \equiv |c\lambda + d|\;.
\eea

For slowly varying fields, the effective Lagrangian of the open string
theory is the DBI Lagrangian. For a review of this theory see Ref. \cite{7}
and references therein.
The equivalence of the noncommutative and ordinary DBI theories has been 
proven \cite{8}. We shall concentrate on both of these theories.
 
In section 2, we shall present an $SL(2;R)$ invariant 
argument for the ordinary and noncommutative DBI Lagrangians. 
Therefore, for special $C_{\mu\nu}$ a $Dp$-brane with ordinary worldvolume,
but modified tension will be obtained. In addition, we obtain the auxiliary
$U(1)$ gauge field strength
${\bar F}_{\mu\nu}$ \cite{9} in terms of the other variables.
This field with the $U(1)$ field strength
$F_{\mu\nu}$ form an $SL(2;R)$ doublet.

In section 3, by introducing a consistent relation between $B_{\mu\nu}$
and ${\tilde B}_{\mu\nu}$, a useful rule will be obtained. That is, the
DBI theory and its $SL(2;R)$ dual theory are duals of each other. In other
words, twice dualizing of the DBI theory leaves it invariant.
This reflection also holds for the noncommutative
DBI theory.

In section 4, we shall obtain some relations between the effective
open string variables
and their duals. Thus, $SL(2;R)$ transformations on the noncommutative
DBI Lagrangian can be captured in the tension of the brane. On the
other hand, we have the original noncommutative DBI theory with the modified
tension. This form of the dual theory
leads to another solution for the auxiliary gauge field.

In section 5, the noncommutativity parameter is related
to its $T$-dual and its $SL(2;R)$ dual versions. 
We shall see that on the effective
variables, $T$-duality and $SL(2;R)$ duality do not commute.
In addition, 
the invariance of the quantity $\frac{G_s}{g_s}$, under the
$T$-duality and $SL(2;R)$ duality will be shown.

In section 6, we study the Chern-Simons (CS) action. For its commutative
theory, for example, see Ref.\cite{10} and for its noncommutative
version, e.g. see Ref.\cite{11,12}. The effects of the $SL(2;R)$ group
on the noncommutative CS action will be studied. We observe that
under twice dualization this action remains invariant.
\section{Noncommutative DBI Lagrangian and its $SL(2;R)$ duality}
Now we study the action of the $SL(2;R)$ group on the noncommutative
DBI Lagrangian. We consider an arbitrary $Dp$-brane.
Consider the noncommutative DBI Lagrangian \cite{8}
\bea
{\widehat {\cal{L}}}_{(0)} = \frac{1}{(2\pi)^p
(\alpha')^{\frac{p+1}{2}}
G^{(0)}_s } \sqrt{\det ( G_{(0)}+ 2\pi \alpha' {\widehat F} )}\;,
\eea
where the index zero shows the cases with zero extra modulus,
i.e. $\Phi=0$.
From the definitions of the open string variables $G_{(0)\mu\nu}$, 
$G^{(0)}_s$ and the noncommutativity parameter $\theta^{\mu\nu}_0$,
in terms of the
closed string variables $g_{\mu\nu} , B_{\mu\nu}$ and $g_s$
(whit $\mu,\nu =0,1,...,p$),
\bea
&~&G_{(0)\mu\nu}=g_{\mu\nu}-(2\pi \alpha')^2 (Bg^{-1}B)_{\mu\nu},
\nonumber\\
&~&\theta^{\mu\nu}_0=-(2\pi \alpha')^2 \bigg{(}
\frac{1}{g+2\pi \alpha'B}B\frac{1}{g-2\pi \alpha'B} \bigg{)}^{\mu\nu},
\nonumber\\
&~&G^{(0)}_s = g_s \bigg{(}\frac{\det(g+2\pi \alpha'B)}{\det g}
\bigg{)}^{1/2},
\eea
one can find their $SL(2;R)$ transformations.
We also require transformation of ${\widehat F}_{\mu\nu}$.

According to the following relation \cite{8}
\bea
{\widehat F} = (1 + F\theta)^{-1}F\;,
\eea
transformation of the noncommutative field strength ${\widehat F}_{\mu\nu}$ 
can be obtained from the transformations of $\theta^{\mu\nu}$ and
the ordinary field strength $F_{\mu\nu}$.

It has been discussed by Townsend \cite{9} that
for $D$-string there are two $U(1)$ gauge fields $F_{\mu\nu}$
and ${\bar F_{\mu\nu}}$, which form an $SL(2;R)$ doublet,
related to the doublet
$\left( \begin{array}{c} B_{\mu\nu} \\  C_{\mu\nu} 
\end{array} \right)$. 
Also see the Ref.\cite{13}. 
We assume that the field strength ${\bar F}_{\mu\nu}$ can be applied to
any $Dp$-brane. Therefore, $F_{\mu\nu}$
and ${\bar F_{\mu\nu}}$ can be interpreted as DBI fields, but not 
both simultaneously. 
Thus, the ordinary gauge field strengths $F_{\mu\nu}$ and
${\bar F}_{\mu\nu}$ 
transform in the same way as the fields 
$B_{\mu\nu}$ and $C_{\mu\nu}$, 
\bea
&~& F_{\mu\nu} \rightarrow {\tilde F}_{\mu\nu} = dF_{\mu\nu}
-c{\bar F}_{\mu\nu}\;,
\nonumber\\
&~& {\bar F}_{\mu\nu} \rightarrow {\widetilde {\bar F}}_{\mu\nu} = 
-bF_{\mu\nu}+a{\bar F}_{\mu\nu}\;.
\eea
Imposing the $SL(2;R)$ 
invariance on the ordinary (noncommutative) DBI theory, gives
${\bar F}_{\mu\nu}$ in terms of $F_{\mu\nu}$ ($F_{\mu\nu}$ and 
$\theta^{\mu\nu}$).
\subsection{Commutative result}

Consider the case $C_{\mu\nu}=\frac{d}{c}B_{\mu\nu}$.
This means that the field ${\tilde B}_{\mu\nu}$ is zero. In other words,
the transformed theory is not noncommutative. Therefore,
the $SL(2;R)$ transformation of the Lagrangian (3) reduces to 
\bea
{\tilde {\cal{L}}} = \frac{1}{(2\pi)^p (\alpha')^{\frac{p+1}{2}}\eta^2 g_s} 
\sqrt{\det \bigg{(} \eta g + 2\pi \alpha' (dF-c{\bar F}) \bigg{)}}\;.
\eea
For ${\bar F}=\frac{d-\eta}{c}F-\frac{\eta}{c}B$, the 
Lagrangian (7) is proportional to the DBI Lagrangian, i.e.,
\bea
{\tilde {\cal{L}}}= \eta^{\frac{p-3}{2}}{\cal{L}}_{DBI}.
\eea 
This equation can be interpreted as follows.
The Lagrangian ${\tilde {\cal{L}}}$ describes the same
$Dp$-brane which is described by ${\cal{L}}_{DBI}$, but with the modified
tension
\bea
{\bar T}_p=\frac{\eta^{(p-3)/2}}{(2\pi)^p(\alpha')^{(p+1)/2}g_s}.
\eea
For the $D3$-brane, theory is symmetric, i.e. ${\tilde {\cal{L}}}=
{\cal{L}}_{DBI}$, as expected. For the strong coupling of strings
$g_s \rightarrow \infty$, the modified tension ${\bar T}_p$ goes to zero.
For the weak coupling constant $g_s \rightarrow 0$, this tension for
$D$-particle goes to zero, for $D$-string is finite and for 
$Dp$-brane with $p \geq 2$ approaches to infinity.
We also can write
\bea
{\bar T}_p = \frac{1}{(2\pi)^p (\tilde \alpha')^{\frac{p+1}{2}}{\tilde g}_s}
\;\;,\;\;\;\;
{\tilde \alpha'} = \frac{\alpha'}{\eta}.
\eea
\subsection{Noncommutative result with $\Phi =0$}

Now we find the conditions for the invariance of the noncommutative theory
with zero modulus $\Phi$. Consider
the following relations between the scalars and 2-forms 
\bea
&~&e^{2\phi} = \frac{c^2}{1-(c\chi+d)^2}\;,
\nonumber\\
&~&C_{\mu\nu} = \frac{d-1}{c}B_{\mu\nu}\;,
\eea
which are equivalent to $\eta =1$ and ${\tilde B}_{\mu\nu}=B_{\mu\nu}$,
respectively.
These assumptions lead to the relations 
${\tilde G}^{(0)}_s = G^{(0)}_s$, 
${\tilde G}_{(0)\mu\nu}=G_{(0)\mu\nu}$ and
${\tilde \theta}^{\mu\nu}_0 = \theta^{\mu\nu}_0$. In addition, 
the field strength should be selfdual or anti-selfdual, i.e., 
${\widetilde {\widehat F}}= \pm {\widehat F}$.
Therefore, the noncommutative theory (3) becomes
$SL(2;R)$ invariant. Since for any matrix $M$ there is $\det M = \det M^T$,
the anti-selfdual case for the Lagrangian (3) also is available.
In other words, we have
$\det (G_{(0)}-2\pi \alpha' {\widehat F})=
\det (G_{(0)}-2\pi \alpha' {\widehat F})^T=
\det (G_{(0)}+2\pi \alpha' {\widehat F})$.

According to the equation (5), the condition on the field strength 
${\widehat F}$ gives ${\tilde F}$ and consequently
${\bar F}$ in terms of $F$ and $\theta_0$,
\bea
{\bar F} = \frac{d}{c}F \mp \frac{1}{c}[F^{-1}+(1\mp 1)\theta_0]^{-1}\;.
\eea
This is a way to determine the auxiliary field strength ${\bar F}_{\mu\nu}$.
For the selfdual case (i.e. the upper signs) the field
strength ${\bar F}$ is proportional to $F$.
For the anti-selfdual case (i.e. the lower signs) we have
\bea
\frac{1}{F}+\frac{1}{{\tilde F}}=\frac{2}{-\theta_0^{-1}}\;.
\eea
This means that, ${-\theta_0^{-1}}$ is 
harmonic mean between $F$ and ${\tilde F}$.
The equation (13) for the commutative case gives an anti-selfdual $F$, i.e.
${\tilde F}=-F$.
\subsection{Noncommutative result including $\Phi$}

The noncommutative DBI Lagrangian with arbitrary noncommutativity parameter
has the dual form
\bea
{\widetilde {\widehat {\cal{L}}}} = \frac{1}{(2\pi)^p
(\alpha')^{\frac{p+1}{2}} {\tilde G}_s }
\sqrt{\det \bigg{(} {\tilde G}+ 2\pi \alpha' ({\tilde \Phi}+
{\widetilde {\widehat F}}) \bigg{)}}\;,
\eea
where the effective parameters ${\tilde G}$, ${\tilde \Phi}$ 
and ${\tilde G}_s$ have been given by the equations
\bea
\frac{1}{{\tilde G}+2\pi \alpha' {\tilde \Phi}} = - \frac{{\tilde \theta}}
{2\pi \alpha'}+\frac{1}{{\tilde g}+2\pi \alpha' {\tilde B}}\;,
\eea
\bea
{\tilde G}_s = \frac{{\tilde g}_s} 
{\sqrt{\det \bigg{(} 1- \frac{{\tilde \theta}}
{2\pi \alpha'}({\tilde g}+2\pi\alpha'{\tilde B})\bigg{)}}}\;.
\eea

For the equations (11) and selfdual $\theta$,  
the dual Lagrangian (14) is equal to the noncommutative Lagrangian
${\widehat {\cal{L}}}$ 
(i.e., equation (14) without tildes) if ${\widehat F}$ is selfdual or
\bea
{\widetilde{\widehat F}}=-{\widehat F}-2\Phi\;.
\eea
To show invariance under this condition, again use the identity
$\det M = \det M^T$. 

According to the equation (17) and dual form of the equation (5),
the field strength ${\bar F}$ is
\bea
\bar F = \frac{d}{c}F+\frac{1}{c} \omega(1+\theta\omega)^{-1}\;,
\eea
where the matrix $\omega$ is
\bea
\omega= (1+F\theta)^{-1}\bigg{(}F(1+2\theta\Phi)+2\Phi \bigg{)}\;.
\eea
As expected, the equation (18) for $\Phi=0$ reduces to the 
equation (12) with plus signs.
\section{Duality of the dual theories}

Define the matrix ${\tilde \Lambda}$ as 
\bea
{\tilde \Lambda} \equiv \left( \begin{array}{cc} {\tilde a} & {\tilde b}\\ 
{\tilde c} & {\tilde d}
\end{array} \right)  
= \left( \begin{array}{cc}
d & -b\\
-c & a
\end{array} \right) = \Lambda^{-1}\;.
\eea
Therefore, we can write
\bea
\lambda = \frac{{\tilde a} {\tilde \lambda}+{\tilde b}}{{\tilde c} 
{\tilde \lambda}+{\tilde d}}\;\;\;\;,\;\;\;\;
\left( \begin{array}{c} B_{\mu\nu} \\ C_{\mu\nu} \end{array} \right)
= ({\tilde \Lambda}^T)^{-1}
\left( \begin{array}{c}
{\tilde B}_{\mu\nu} \\ {\tilde C}_{\mu\nu} 
\end{array} \right)\;.  
\eea
Also let the parameter ${\tilde \eta}$ be 
\bea
{\tilde \eta} \equiv | {\tilde c}{\tilde \lambda}+{\tilde d}|
= \frac{1}{\eta}\;.
\eea
This gives
\bea
g_s = {\tilde \eta}^2{\tilde g}_s\;,\;\;\;g_{\mu\nu} = {\tilde \eta}
{\tilde g}_{\mu\nu}.
\eea
That is, in some equations if we change the dual
quantities with the initial quantities the resulted equations
also hold. 
With this rule, the equations (21) and (23) directly can be obtained
from the equations (1) and (2).

For generalization of the above rule let the 2-form 
$C_{\mu\nu}$ be proportional to $B_{\mu\nu}$ as in the following
\bea
C_{\mu\nu}=\frac{d-\eta}{c}B_{\mu\nu}\;.
\eea
This leads to the relation
\bea
{\tilde B}_{\mu\nu}=\eta B_{\mu\nu}\;,
\eea
or equivalently ${\tilde C}_{\mu\nu}= \frac{1-a\eta}{d-\eta}C_{\mu\nu}$.
These equations also hold under the exchange of the dual quantities with the
initial quantities. In other words, we have 
$({\tilde d}-{\tilde \eta}){\tilde B}_{\mu\nu}
={\tilde c}{\tilde C}_{\mu\nu}$,  
$B_{\mu\nu}={\tilde \eta} {\tilde B}_{\mu\nu}$ and
$C_{\mu\nu}= \frac{1-{\tilde a}{\tilde \eta}}{{\tilde d}-{\tilde \eta}}
{\tilde C}_{\mu\nu}$.

According to the equations (2), (4) and (25), for the zero modulus
$\Phi$, the transformations of $G_{(0)}$, $\theta_0$ and $G^{(0)}_s$ are
as in the following
\bea
&~&{\tilde G}_{(0)\mu\nu}=\eta G_{(0)\mu\nu},
\nonumber\\
&~&{\tilde \theta}^{\mu\nu}_0=\frac{{\theta}^{\mu\nu}_0}{\eta},
\nonumber\\
&~&{\tilde G}^{(0)}_s=\eta^2 G^{(0)}_s.
\eea
On the other hand, these equations also obey from the above rule.

Since ${\tilde \Lambda}\in SL(2;R)$, we conclude that
the initial theory also is $SL(2;R)$ transformed of the dual theory. 
Therefore, the mentioned rule can be written as
\bea
\;\;\;\;\;\;\;\;\;\;\;\;\;\;\;
{\rm Initial\;theory} \stackrel{\Lambda}\longrightarrow 
SL(2;R){\rm\;dual\;theory},
\nonumber\\
\;\;\;\;\;\;\;\;\;\;\;\;\;\;\;
SL(2;R){\rm\;dual\;theory} \stackrel{\tilde \Lambda}\longrightarrow
{\rm Initial\;theory}.  
\eea
In other words, twice dualization leaves the theory (and related equations)
invariant.
Note that ``{\it initial theory}'' refers to the type IIB theory or DBI theory.
In the next sections, we shall see that the rule (27) will be repeated.
For example, it also holds for the noncommutative DBI theory, ordinary
and noncommutative Chern-Simons actions.
The statement (27) for the ordinary DBI theory is obvious, i.e.,
\bea
{\tilde{\tilde {\cal{L}}}}_{DBI} = {\cal{L}}_{DBI}.
\eea
\section{Relations between the effective variables}
The noncommutative DBI Lagrangian and the $SL(2;R)$ 
duality of it can be described more 
generally, such that the noncommutativity parameters $\theta$ and 
${\tilde \theta}$ become arbitrary \cite{8}. Therefore, the extra moduli
$\Phi$ and ${\tilde \Phi}$ are not zero (for example, the dual theory
was given by the equation (14)). 
The equations (26) guide us to
introduce the following relations between the
effective metrics and the extra moduli
\bea  
&~&{\tilde G}_{\mu\nu}=\eta G_{\mu\nu}\;,
\nonumber\\
&~&{\tilde \Phi}_{\mu\nu}=\eta \Phi_{\mu\nu}\;.
\eea
According to the equations (15) and (29) we obtain
\bea
{\tilde \theta}^{\mu\nu}=\frac{\theta^{\mu\nu}}{\eta}\;.
\eea
This implies that, 
if the effective theory is noncommutative (ordinary) the dual
theory of it also is noncommutative (ordinary). 
Note that if we introduce the equation (30) then we 
obtain the equations (29).
The equations (16) and (30) give the following relation between the effective string
couplings ${\tilde G}_s$ and $G_s$,
\bea
{\tilde G}_s=\eta^2 G_s\;.
\eea

The equations (29)-(31) have the following properties. (a)
They are consistent
with the rule (27). In other words, they can be written in the forms
$G_{\mu\nu}={\tilde \eta} {\tilde G}_{\mu\nu}$, 
$\Phi_{\mu\nu}={\tilde \eta} {\tilde \Phi}_{\mu\nu}$, 
$\theta^{\mu\nu}=\frac{{\tilde \theta}^{\mu\nu}}{{\tilde \eta}}\;$ and 
$G_s={\tilde \eta}^2 {\tilde G}_s$. (b) For the
commutative case, i.e., $\theta=0$ we have ${\tilde \theta}=0$. Thus, 
the equations (29) change to ${\tilde g_{\mu\nu}}=\eta g_{\mu\nu}$ and 
${\tilde B_{\mu\nu}}=\eta B_{\mu\nu}$. (c) For the variables
\bea
&~&\theta=B^{-1}\;\;,\;\;G=-(2\pi \alpha')^2Bg^{-1}B\;\;,\;\; \Phi=-B\;,
\nonumber\\
&~&{\tilde \theta}={\tilde B}^{-1}\;\;,\;\;{\tilde G}=-(2\pi \alpha')^2 
{\tilde B}{\tilde g}^{-1}{\tilde B}\;\;,\;\; {\tilde \Phi}=-{\tilde B}\;,
\eea
the equations (29) and (30) reduce to identities. Note that these 
variables also satisfy the equation (15) and this equation without tildes.
(d) For $\theta =\theta_0$ the equation (30) gives 
${\tilde \theta} ={\tilde \theta}_0$, therefore, $\Phi={\tilde \Phi}=0$.
In this case as expected, there are $G=G_{(0)}$, ${\tilde G}=
{\tilde G}_{(0)}$, $G_s=G_s^{(0)}$ and ${\tilde G}_s 
={\tilde G}_s^{(0)}$. On the other hand, the equations (29)-(31) reduce to
the results (26).

The equations (25), (29), (30) and the second equation of (2)
lead to the relations
\bea
&~&{\tilde Q_{(1)}}^{\mu\nu}{\tilde Q}_{(2)\rho\sigma}= 
Q_{(1)}^{\mu\nu}Q_{(2)\rho\sigma}\;,
\nonumber\\
&~&{\tilde Q}^{\mu\nu}_{(1)}Q_{(2)\rho\sigma}= 
Q^{\mu\nu}_{(1)}{\tilde Q}_{(2)\rho\sigma}\;,
\eea
where $Q_1, Q_2 \in \{G,\Phi, \theta,g,B\}$. That is, the quantities 
$Q_{(1)}^{\mu\nu}Q_{(2)\rho\sigma}$ and ${\tilde Q}^{\mu\nu}_{(1)} 
Q_{(2)\rho\sigma}$ are $SL(2;R)$ invariant. On the other hand, since we
have ${\tilde {\tilde Q}}_{(i)}=Q_{(i)}$ for $i = 1,2$, the equations
(33) are consistent with the rule (27).

According to the equations (29) and (31), for the following values of
the dual field ${\widetilde {\widehat F}}$,
\bea
{\widetilde {\widehat F}} = \eta [\pm {\widehat F}-(1\mp 1)\Phi]\;,
\eea
the dual Lagrangian (14) takes the form 
\bea
{\widetilde {\widehat {\cal {L}}}}=
\eta^{\frac{p-3}{2}}{\widehat {\cal {L}}}.
\eea
After substituding (34) in (14), for the lower signs one should perform
transpose on the matrices in (14). Again use the identity 
$\det M =\det M^T$. This Lagrangian describes
the same noncommutative $Dp$-brane, which also is given by 
${\widehat {\cal {L}}}$, but with the modified tension, i.e., 
${\bar {\widehat T}}_p=\eta^{\frac{p-3}{2}}{\widehat T}_p $.
For the $D3$-brane the theory is invariant. For the commutative case, the
equation (35) reduces to the equation (8), as expected.

The equation (34) with lower signs, is generalization 
of the equation (17). The field strength ${\bar F}$, extracted from the 
equation (34), is
\bea
{\bar F} = \frac{d}{c}F+\frac{1}{c} \Omega(1+\frac{1}{\eta}\theta 
\Omega)^{-1}\;,
\eea
where the matrix $\Omega$ is
\bea
\Omega = \eta(1+F\theta)^{-1} \bigg{(} F[\mp 1 + (1 \mp 1)\theta \Phi ]
+(1 \mp 1)\Phi \bigg{)}\;.
\eea

Since the equation (34) can be written as 
${\widehat F} = {\tilde \eta} [\pm {\widetilde{\widehat F}}-
(1\mp 1){\tilde \Phi}]$, this with the equations (29)-(31) imply that, in the
rule (27) the ``{\it initial theory}'' also can be the noncommutative
DBI theory. This also can be seen from the equation (35), i.e.
${\widehat {\cal {L}}}=
{\tilde \eta}^{\frac{p-3}{2}}{\widetilde {\widehat {\cal {L}}}}$,
or equivalently
\bea
{\widetilde{\widetilde {\widehat{\cal{L}}}}} = {\widehat{\cal{L}}}.
\eea
\section{Commutator of the $SL(2;R)$ duality and $T$-duality
on the effective variables}

Previously we have observed that the effective metric $G_{(0)\mu\nu}$ and the
noncommutativity parameter $\theta^{\mu\nu}_0$ under 
the $T$-duality transform
to $G'_{(0)\mu\nu}$ and ${\theta'}^{\mu\nu}_0$ as in
the following \cite{14}
\bea
&~&G'_{(0)\mu\nu}=g_{\mu\nu}\;,
\nonumber\\
&~&{\theta'_0}^{\mu\nu}=(2\pi \alpha')^2 (B^{-1})^{\mu\nu}\;.
\eea
The actions of $SL(2;R)$ duality on these equations give
\bea
&~&\widetilde {(G'_0)}_{\mu\nu} =\eta g_{\mu\nu}\;,
\nonumber\\
&~&\widetilde {(\theta'_0)}^{\mu\nu} = 
\frac{1}{\eta}(2\pi \alpha')^2 (B^{-1})^{\mu\nu}\;.
\eea
Application of $T$-duality on the first and second equations of (26)
and then comparison
of the results with the equations (40), lead to the relations
\bea
&~&[\widetilde {(G'_0)} -({\tilde G}_0)']_{\mu\nu}=
(\eta-\eta')G'_{(0)\mu\nu} \;,
\nonumber\\
&~&[\widetilde {(\theta'_0)} -({\tilde \theta}_0)']^{\mu\nu}=
\bigg{(}\frac{1}{\eta}-\frac{1}{\eta'}\bigg{)} {\theta'_0}^{\mu\nu}\;,
\eea
where $\eta'$ is $T$-duality of $\eta$.
These equations imply that, on the open string metric and noncommutativity 
parameter, unless $\eta=\eta'$, $T$-duality and $SL(2;R)$ duality do 
not commute with each other. We shall show that for the nonzero modulus 
$\Phi$, these equations also hold.

In the presence of the extra modulus $\Phi$, we have the following 
relation \cite{14} 
\bea
G'+2\pi \alpha' \Phi'= (g+2\pi \alpha' B)^{-1}
(G-2\pi \alpha' \Phi) (g-2\pi \alpha' B)^{-1}\;.
\eea
Therefore, there is the following relation between the 
noncommutativity parameter $\theta^{\mu\nu}$ and its $T$-duality 
${\theta'}^{\mu\nu}$, 
\bea
{\theta'}^{\mu\nu}=-[(g-2\pi \alpha'B) \theta (g+2\pi\alpha'B)
]^{\mu\nu}\;.
\eea
According to this equation and equation (30) we obtain
\bea
{\theta'}^{\mu\nu}=-\eta [(g-2\pi \alpha'B) {\tilde \theta} 
(g+2\pi\alpha'B)]^{\mu\nu}\;.
\eea
That is, the $T$-duality and $SL(2;R)$ duality versions 
of the noncommutativity parameter are related to each other. 

Action of the $SL(2;R)$ duality on $G'$, $\Phi'$ and $\theta'$ of the 
equations (42) and (43) and also action of $T$-duality on 
${\widetilde G}$, ${\widetilde \Phi}$ and ${\widetilde \theta}$ of the 
equations (29) and (30), and then comparison of the results, give
\bea
&~&[\widetilde {(Q')}-({\tilde Q})']_{\mu\nu}=(\eta-\eta') Q'_{\mu\nu}\;,
\nonumber\\
&~&[\widetilde {(\theta')}-({\tilde \theta})']^{\mu\nu}
=(\frac{1}{\eta}-\frac{1}{\eta'}) {\theta'}^{\mu\nu}\;,
\eea
where $Q\in \{G,\Phi \}$. Let us denote the dualities of $Q$ as
$Q' \equiv TQ $ and ${\tilde Q} \equiv SQ$. Thus, the equations (45) take
the forms
\bea
&~&([S,T] Q)_{\mu\nu}=(\eta-\eta') (TQ)_{\mu\nu}\;,
\nonumber\\
&~&([S,T]\theta)^{\mu\nu}=(\frac{1}{\eta}-\frac{1}{\eta'})
(T\theta)^{\mu\nu}\;.
\eea
Similarly, for the effective string coupling $G_s$ there is
\bea
[S,T] G_s=(\eta^{(3-p)/2}-\eta'^2) (TG_s).
\eea
Therefore, on the variables $G$, $\Phi$, $\theta$ and $G_s$
$T$-duality and $SL(2;R)$ duality do not commute.
In other words, the commutator of 
these dualities, is proportional to the effects of $T$-duality.

The $T$-duality of the effective string coupling is $G'_s=\frac{G_s}{
\sqrt{ \det (g+2\pi \alpha'B)}}$. This implies $\frac{G_s}{g_s}$ is a 
$T$-duality invariant quantity \cite{14}.
From this and the equation (31) we conclude that
\bea
\frac{G'_s}{g'_s}=\frac{{\tilde G}_s}{{\tilde g}_s}=\frac{G_s}{g_s}\;.
\eea
That is, the ratio $\frac{G_s}{g_s}$ also is invariant under the 
$SL(2;R)$ duality. Therefore, on the quantity $\frac{G_s}{g_s}$,
$T$-duality and $SL(2;R)$ duality commute.
\section{$SL(2;R)$ duality of the noncommutative Chern-Simons action}

The DBI action describes the couplings of a $Dp$-brane to the massless
Neveu-Schwarz fields $g_{\mu\nu}$, $B_{\mu\nu}$ and $\phi$. The interactions
with the massless Ramond-Ramond (R-R) fields are incorporated in the Chern-
Simons action \cite{10}
\bea
S_{CS}=\frac{1}{(2\pi)^p(\alpha')^{(p+1)/2}g_s} \int \sum_n
C^{(n)} \wedge e^{2\pi\alpha'(B+F)}\;,
\eea
where $C^{(n)}$ denotes the $n$-form R-R potential. The exponential should
be expanded so that the total forms have the rank of the worldvolume
of brane. In fact, this action is for a single BPS $Dp$-brane.

The noncommutative Chern-Simons action for constant fields can be written
as in the following \cite{11}
\bea
{\widehat S}_{CS}=\frac{1}{(2\pi)^p(\alpha')^{(p+1)/2}g_s} \int
\sqrt{\det (1-\theta {\widehat F})}\sum_n
C^{(n)} \wedge \exp \bigg{(} 2\pi\alpha'[B+{\widehat F}(1-\theta
{\widehat F})^{-1}] \bigg{)}\;,
\eea
also see Ref.\cite{12}. This action holds for general modulus $\Phi$. It describes the
R-R couplings to a noncommutative $Dp$-brane.

Now we study the effects of the $SL(2;R)$ group on this action. We can apply
${\widetilde {\widehat F}}$ from (34). For simplicity, choose the upper
signs for ${\widetilde {\widehat F}}$. In addition, the equations (25)
and (30) can be used for ${\tilde B}$ and ${\tilde \theta}$. Adding all
these together, we obtain
\bea
{\widetilde {\widehat S}}_{CS}=\frac{1}{(2\pi)^p(\alpha')^{(p+1)/2}\eta^2 g_s}
\int \sqrt{\det (1-\theta {\widehat F})}\sum_n
{\tilde C}^{(n)} \wedge \exp \bigg{(} 2\pi\alpha'\eta [B+{\widehat F}(1-\theta
{\widehat F})^{-1}] \bigg{)}\;.
\eea
Therefore, we should determine the dual fields $\{{\tilde C}^{(n)}\}$.
Since our attention is on the type IIB theory, ${\tilde C}^{(n)}$ is an
even form. The dual fields ${\tilde C}^{(0)} \equiv {\tilde \chi}$ and
${\tilde C}^{(2)}\equiv {\tilde C}$ have been given by the transformations
(1). The field $C^{(4)}$ corresponds to the $D3$-brane. It was
shown in \cite{3,4} that the invariance of the equations of motion,
extracted from the total action $S_{DBI}+S_{CS}$, under the $SL(2;R)$
group, gives the transformations (1) and
\bea
C^{(4)} \rightarrow {\tilde C}^{(4)}=C^{(4)}.
\eea
For the forms ${\tilde C}^{(6)}$, ${\tilde C}^{(8)}$ and
${\tilde C}^{(10)}$ one may use the Hodge duals of the forms
${\tilde C}^{(4)}$, ${\tilde C}^{(2)}$ and ${\tilde C}^{(0)}$,
which are available. However, we have the following results at least for
$n \leq 4$.

The noncommutative Chern-Simons action (50) respects the rule (27), if
twice dualization of the R-R fields are invariant
\bea
{\widetilde {\widetilde C}}^{(n)}=C^{(n)}.
\eea
From the transformations (1) and (52) explicitly one can see this equation
for $C^{(0)}$, $C^{(2)}$ and $C^{(4)}$.
On the other hand, using (53) (at least for $n\leq 4$) 
and then applying $SL(2;R)$ transformations on
the dual action (51), we obtain
\bea
{\widetilde {\widetilde {\widehat S}}}_{CS}={\widehat S}_{CS}.
\eea

From the equations (1), (2), (6) and (20) we have
${\tilde {\tilde B}}_{\mu\nu}= B_{\mu\nu}$,
${\tilde {\tilde g}}_s= g_s$ and ${\tilde {\tilde F}}_{\mu\nu}= F_{\mu\nu}$.
By considering the equation (53), we observe that the ordinary
Chern-Simons action (49) also obey the rule (27),
\bea
{\widetilde {\widetilde S}}_{CS}=S_{CS}.
\eea
For vanishing noncommutativity parameter, the equation (54) reduces to
(55), as expected.
\section{Conclusions} 
We studied the action of the $SL(2;R)$ group on the
noncommutative DBI theory with zero and nonzero extra modulus $\Phi$.
The invariance of the theory determines
the corresponding noncommutative field strength
${\widehat F}_{\mu\nu}$. As a consequence, the
auxiliary field strength ${\bar F}_{\mu\nu}$ has been obtained.
For a special value of the R-R 2-form, 
the $SL(2;R)$ group on the noncommutative DBI 
Lagrangian produces a theory which describes an ordinary brane with
the modified tension. For the $D3$-brane the resulted ordinary theory is 
DBI theory, as expected.

We observed that the extracted equations of the ordinary DBI and
noncommutative DBI theories 
under the exchange of the variables with their dual variables
are invariant.
In other words, twice dualizing of these theories and 
the corresponding
variables and equations, does not change them.
This implies that these theories and their $SL(2;R)$
transformations, are dual of each other.

By introducing some relations
(which are consistent with the rule (27))
between the effective variables and their duals,
we obtained some other 
equations that are $SL(2;R)$ invariant. Therefore, another solution for the
auxiliary gauge field was found.
In this case, $SL(2;R)$ duality of the noncommutative DBI theory is
proportional to the noncommutative DBI theory. For the $D3$-brane the
theory is selfdual.

We showed that the noncommutativity
parameter, its $T$-dual and its $SL(2;R)$ dual have relations with 
each other. We found that on 
the open string metric, noncommutativity parameter, the extra 
modulus $\Phi$ and the effective string coupling,
$T$-duality and $SL(2;R)$ duality do not commute.
We also observed that the ratio of the effective string
coupling to the string coupling under the above dualities is invariant.

Finally, we studied the effects of the $SL(2;R)$ group on the noncommutative
Chern-Simons action. Under two successive dualizations, similar the DBI theory,
this action remains invariant. This also occurs for the ordinary
Chern-Simons action.


\begin{thebibliography}{99} 

\bibitem{1}
S.J. Rey, Phys. Rev. {\bf D43}(1991)526; A. Font, L. Ibanez, D. Lust and 
F. Quevedo, Phys. Lett. {\bf B249}(1990)35;
M.J. Duff and R.R. Khuri, Nucl. Phys. {\bf B411}(1994)473; A. Sen, Nucl.
Phys. {\bf B404}(1993)109; Phys. Lett. {\bf B303}(1993)22; Mod. Phys. Lett.
{\bf A8}(1993)2023; J.H. Schwarz and A. Sen, Phys. Lett. {\bf B312}(1993)105;
M.J. Duff and J. Rahmfeld,
Phys. Lett. {\bf B345}(1995)441.
\bibitem{2}
C. Hull and P.K. Townsend, Nucl. Phys. {\bf B438}(1995)109;
A. Sen, Int. J. Mod. Phys. {\bf A9}(1994)3707, hep-th/9402002.
\bibitem{3}
A.A. Tseytlin, Nucl. Phys. {\bf B469}(1996)51, hep-th/9602064.
\bibitem{4}
M.B. Green and M. Gutperle, Phys. Lett. {\bf B377}(1996)28,
hep-th/9602077.
\bibitem{5}
E. Witten, Nucl. Phys. {\bf B460}(1996)335, hep-th/9510135;
M. Dc Roo, Nucl. Phys. {\bf B255}(1985)515; 
J.H. Schwarz, Phys. Lett. {\bf B360}(1995)13, hep-th/9508143; 
G.W. Gibbons and D.A. Rasheed,
Phys. Lett. {\bf B365}(1996)46, hep-th/9509141;
M. Cederwall and P.K. Townsend, JHEP {\bf 9709}(1997)003, 
hep-th/9709002; A. Sen, Phys. Lett.
{\bf B329}(1994)217, hep-th/9402032.
\bibitem{6}
S.J. Rey and R. von Unge, Phys. Lett. {\bf B499}(2001)215, hep-th/0007089; 
R. Gopakumar, J. Maldacena, S. Minwalla and A. Strominger, 
JHEP {\bf 0006}(2000)036, hep-th/0005048; J.X. Lu, S. Roy and 
H. Singh, JHEP {\bf 0009}(2000)020, hep-th/0006193; 
Nucl. Phys. {\bf B595}(2001)298, hep-th/0007168;
D.S. Berman,
Phys. Lett. {\bf B409}(1997)153, hep-th/9706208;
C. Hofman and E. Verlinde, JHEP {\bf 9812}(1998)010, hep-th/9810116;
C.S. Chan, A. Hashimoto and H. Verlinde, JHEP {\bf 0109}(2001)034, hep-th/0107215;
R.G. Cai and N. Ohta, Prog. Theor. Phys. {\bf 104}(2000)1073,
hep-th/0007106.
\bibitem{7}
M. Born and L. Infeld, Proc. R. Soc. {\bf A144}(1934)425;
R.G. Leigh, Mod. Phys. Lett. {\bf A4}(1989)2767;
A.A. Tseytlin, Nucl. Phys. {\bf B276}(1986)391; Nucl. Phys. {\bf B501}
(1997)41, hep-th/9701125;
E.S. Fradkin and A.A. Tseytlin, Phys. Lett. {\bf B163}(1985)123;
C.G. Callan, C. Lovelace, C.R. Nappi and S.A. Yost,
Nucl. Phys. {\bf B288}(1987)525-550; A. Abouelsaood, C.G. Callan, C.R. Nappi
and S.A. Yost, Nucl. Phys. {\bf B280}(1987)599.
\bibitem{8}
N. Seiberg and E. Witten, JHEP {\bf9909}(1999)032, hep-th/9908142.
\bibitem{9}
P.K. Townsend, Phys. Lett. {\bf B409}(1997)131, hep-th/9705160.
\bibitem{10}
M.R. Douglas, hep-th/9512077;
M. Li, Nucl. Phys. {\bf B460}(1996)351, hep-th/9510161;
M.B. Green, J.A. Harvey and G. Moore, Class. Quant. Grav. {\bf 14}(1997)
47, hep-th/9605033.
\bibitem{11}
S. Mukhi and N.V. Suryanarayana, JHEP {\bf 0011}(2000)006, hep-th/0009101;
hep-th/0107087.
\bibitem{12}
H. Liu and J. Michelson, Phys. Lett. {\bf B518}(2001)143, hep-th/0104139.
\bibitem{13}
P.K. Townsend, Phys. Lett. {\bf B277}(1992)285; E. Bergshoeff,
L.A.J. London and P.K. Townsend, Class. Quant. Grav. {\bf 9} (1992)2545,
hep-th/9206026.
\bibitem{14}
D. Kamani, Mod. Phys. Lett. {\bf A17}(2002)237, hep-th/0107184.

\end{thebibliography}
\end{document}